\documentclass{article}
\usepackage{spconf,amsmath,graphicx,enumitem,pifont}
\usepackage{booktabs,multirow}
\usepackage{arydshln}
\usepackage{balance}

\title{LocSelect: Target Speaker Localization with an Auditory Selective Hearing Mechanism}
\name{Yu Chen$^{1,2}$, Xinyuan Qian$^1$, Zexu Pan$^3$, Kainan Chen$^4$, Haizhou Li$^2$
\thanks{This work has been submitted to the IEEE for possible publication. Copyright may be transferred without notice, after which this version may no longer be accessible}
\thanks{This work is supported by by the Guangdong Provincial Key Laboratory of Big Data Computing, The Chinese University of Hong Kong, Shenzhen (Grant
No. B10120210117-KP02); by the CCF-Tencent Rhino-Bird Open Research Fund; and by National Natural Science Foundation of China, Grant No. 62306029.}
}
\address{
$^1$School of Computer and Communication Engineering, \\University of Science and Technology Beijing (USTB), China\\
$^2$School of Data Science, The Chinese University of Hong Kong, Shenzhen, China \\
 $^3$Institute of Data Science, National University of Singapore, Singapore \\
 $^4$Eigenspace GmbH, Germany}

\begin{document}
\ninept

\maketitle
\begin{abstract}
The prevailing noise-resistant and reverberation-resistant localization algorithms primarily emphasize separating and providing directional output for each speaker in multi-speaker scenarios, without association with the identity of speakers. In this paper, we present a target speaker localization algorithm with a selective hearing mechanism. Given a reference speech of the target speaker, we first produce a speaker-dependent spectrogram mask to eliminate interfering speakers' speech. Subsequently, a Long short-term memory (LSTM) network is employed to extract the target speaker's location from the filtered spectrogram. Experiments validate the superiority of our proposed method over the existing algorithms for different scale invariant signal-to-noise ratios (SNR) conditions. Specifically, at SNR = -10 $dB$, our proposed network LocSelect achieves a mean absolute error (MAE) of $3.55^{\circ}$ and an accuracy 
 (ACC) of 87.40\%.
\end{abstract}


\section{Introduction}

\label{sec:intro}
Speaker localization is primarily concerned with estimating the direction of arrival (DoA) to the microphone array. Traditionally, speaker localization has been approached as a signal-processing problem. The generalized cross-correlation with phase transform (GCC-PHAT) \cite{1162830} and multiple signal classification (MUSIC) \cite{1143830} are the two most popular traditional speaker localization algorithms, both originating from narrowband antenna signal processing. However, the performance of traditional methods is often subpar in challenging multi-speaker scenarios that involve noise, and reverberation. 

Researchers have explored deep neural network (DNN)-based methods to improve multi-speaker localization by leveraging large-scale datasets.
They have successfully combined traditional techniques like GCC-PHAT and MUSIC with DNNs \cite{takeda2018unsupervised,7953333,xue2019direct,wang2018robust,shmuel2023deep,9564893}, making the conventional methods better equipped to handle challenges in multi-speaker scenarios. Other researchers also proposed fully neural network architectures \cite{chakrabarty2017broadband,perotin2018crnn,cao2019polyphonic,nguyen2021general}, showcasing the robustness and efficiency of neural networks 
\begin{figure}[!tb]
    \centering
    \includegraphics[width=1\linewidth]{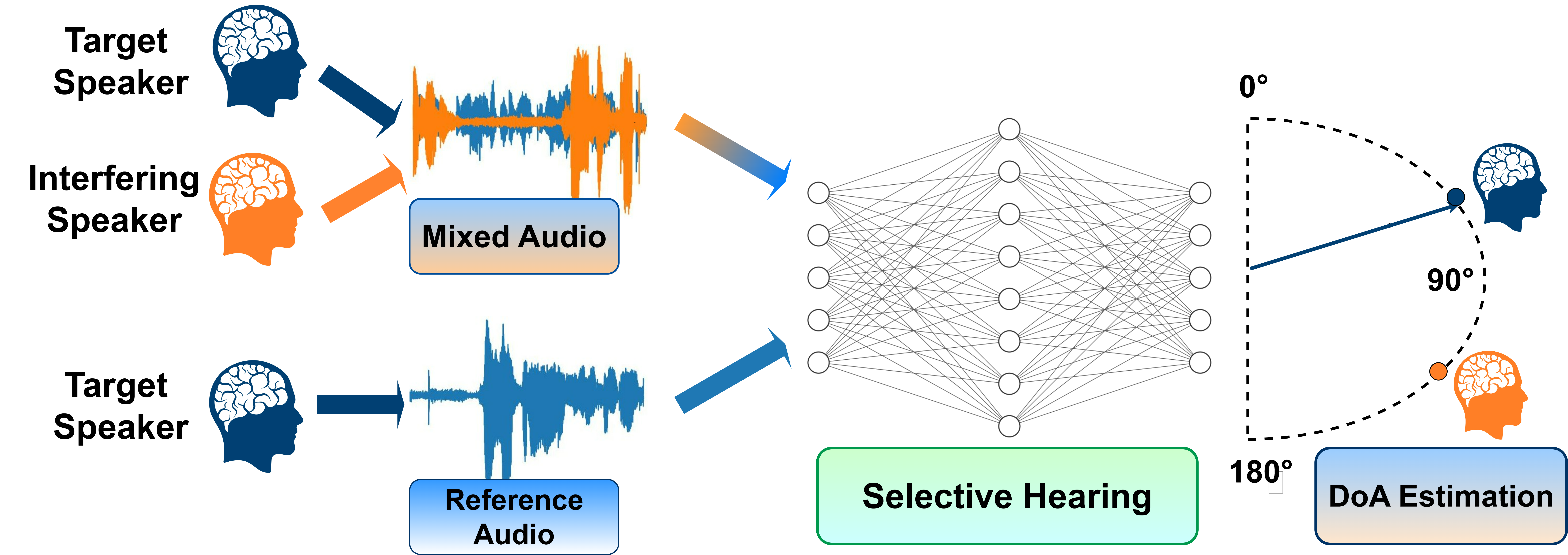}
    \caption{Illustration of our proposed LocSelect:
    Given a reference audio, it is capable of providing the target speaker's DoA  while neglecting the influence from other interfering speakers through the `\textit{selective hearing}' mechanism.}
    \label{fig:my_label}
\end{figure}
in multi-speaker localization. Despite these advancements, a key challenge persists—the association of speaker identities. Existing methods
can  estimate DoA for all speakers, but they struggle to identify the specific target speaker of interest. This limitation hampers practical applications where isolating the target speaker is crucial.

\begin{figure*}[!htb]
\begin{center}
\includegraphics[scale=0.12]{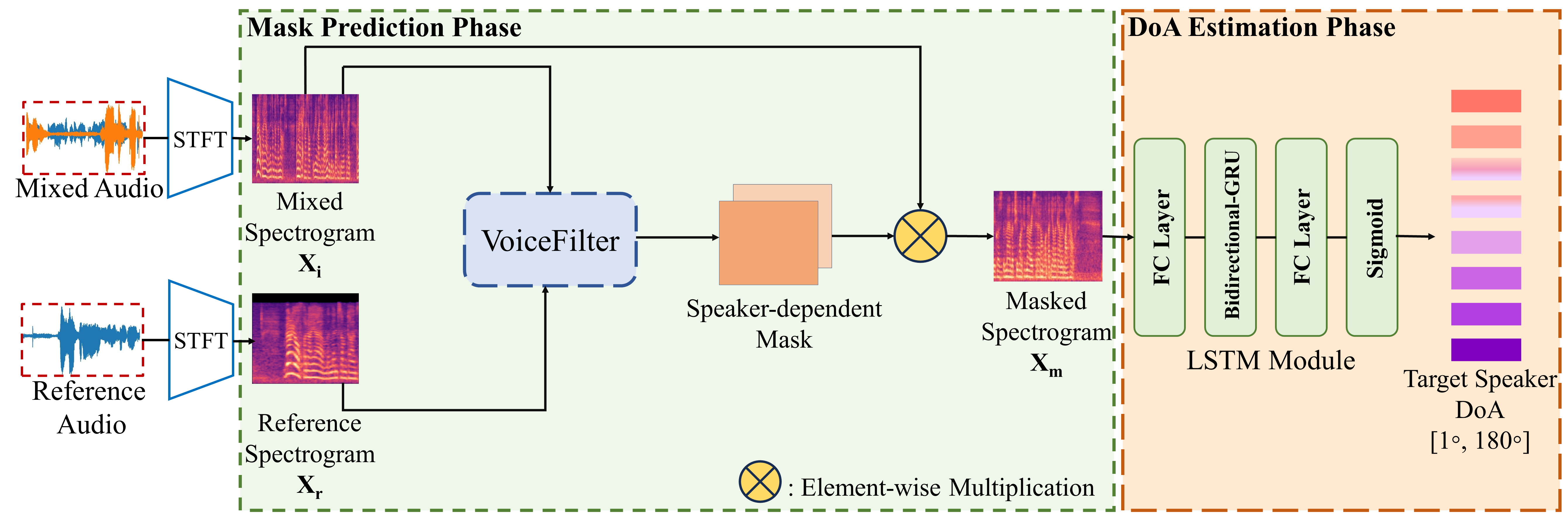}
\end{center}
   \caption{The architecture of our proposed LocSelect for target speaker DoA estimation. It comprises two distinct phases: (1) \textbf{Mask Prediction Phase} (left green block): 
   the spectrograms extracted from the mixed and reference audios are fed into the VoiceFilter, resulting in a speaker-dependent mask which is applied to filter out information from interfering speakers. (2) \textbf{DoA Estimation Phase} (right orange block): The masked spectrogram is fed into a specifically designed LSTM module which only predicts the DoA of the target speaker.}
\label{fig:short}
\end{figure*}

Several studies \cite{sivasankaran2018keyword,wang2018target} tackle the target speaker localization problem. These methods typically begin by estimating a speaker-dependent mask based on specific cues, which is then used in later localization algorithms to deduce DoA of the target speaker. For instance, a reference keyword~\cite{sivasankaran2018keyword} has been employed to obtain a speaker-dependent mask, aiding in characterizing the target speaker. Additionally, in \cite{wang2018target}, an external microphone placed close to the target speaker is used to statistically derive the speaker-dependent mask, combined with the original microphone array signals. However, in real-world settings, cues such as keywords or external microphones may not always be available, posing a limitation to the practicality of these methods.

In practical scenarios, human auditory perception showcases a remarkable cognitive capability, allowing individuals to consciously and effectively direct their attention, especially when they are familiar with the speaker's voice. This auditory acumen equips individuals with the capacity to adeptly filter out ambient noise and distractions, enabling them to focus on and seamlessly track a specific speaker amidst a cacophony of surrounding sounds \cite{4490206}. In this paper, drawing inspiration from human auditory perception, we present a novel target speaker localization approach, namely LocSelect, which integrates the selective hearing mechanism into the LSTM network. Our proposal stands apart from previous target speaker localization works by directly employing a reference speech as a cue to extract the target speaker's location. Specifically, we adapt the VoiceFilter \cite{wang2018voicefilter} to generate the speaker-independent mask, resulting in a filtered spectrogram for precise target speaker localization.

Our experiments validate the effectiveness and robustness of our proposed algorithm, showcasing its potential for real-world applications.

\section{Problem Formulation}

Given the audio signals captured by a microphone pair, we aim to estimate the DoA of the target speaker, denoted as $\theta_{t} \in [1^{\circ},360^{\circ}]$.

We formulate the problem as a regression task, which seeks to predict the DoA as one of the 360 discrete classes, denoted as $\theta$ = \{j $\mid$ j is an integer, 1 $\leq$ j $\leq$ 360 \}.

As DoAs are spatially continuous, instead of adopting the one-hot output coding, we use a Gaussian-like vector \cite{he2018deep}, denoted as $p_{t}(\theta)$, to represent the posterior probability likelihoods of a speaker presence in the direction of $\theta_{t}$,
\begin{equation}
    p_{t}(\theta)=exp(-\frac{\|\theta-\theta_{t}\|}{\sigma_{\theta}})
    \label{eq:posterior}
\end{equation}
where $p_{t}(\theta)$ is centered on the ground truth $\theta_{t}$ with a standard deviation $\sigma_{\theta}$.The prefix $\frac{1}{\sqrt{2\pi\sigma_{\theta}}}$ is dropped since it doesn't influence the result of the algorithm.

Let us denote the Short-time Fourier Transform (STFT) of the audio signal as,
\begin{equation}
    y(t,f) = s(t,f) + \sum_{k=1}^{K} a_{k}(t,f) + n(t,f)
    \label{eq:STFT}
\end{equation}

where $y(t,f) = [y_{1}(t,f), y_{2}(t,f)]^{T}$ denotes the microphone observation vector (2 channel); $t = 1,..., T$ and $f = 1,..., F$ denotes the time and frequency indices; $s(t,f)$, $a_{k}(t,f)$, $n(t,f)$ denote the signal from target speaker, interfering speakers and environmental noise, respectively. Additionally, we denote the clean reference audio as $r(t,f)$.

We adopt a deep neural network to learn the mapping from the STFT inputs to target speaker DoA posterior probability,
\begin{equation}
    \Bar{p}_{t}(\theta) = \mathcal{F}(y(t,f),r(t,f),\Omega)
    \label{eq:neural}
\end{equation}
where $\mathcal{F}(\cdot)$ is the proposed network with $\Omega$ learnable parameters. In this way, the focus of the speaker is approximated by the DoA value with the highest probability,
\begin{equation}
\Bar{\theta}_{t} = \underset{\forall \theta}{\arg\max} \ \Bar{p}_{t}(\theta)
\label{eq:DoA}
\end{equation}

\section{Network Architecture}
The architecture of our novel target speaker localization algorithm, namely LocSelect, is depicted in Fig.2. The proposed network is structured into two distinct phases: the \textit{mask prediction phase} (left green block) and the \textit{DoA estimation phase} (right orange block).

The STFT is a commonly employed technique for extracting audio features, consciously preserving the complete spectral information of the signal. This empowers the network to instinctively identify crucial details for localization, encompassing variations in sound level and phase across channels, alongside the distinct frequency traits of each channel \cite{he2021neural,wang2020ustc,nguyen2020ensemble}. We adopted this method to derive the real and imaginary components of the time-frequency domain signal. Specifically, the magnitude component of this signal is utilized as input during the mask prediction phase, whereas both the masked magnitude component and the phase component are employed in the subsequent DoA estimation phase.

\subsection{Mask Prediction Phase}
In the mask prediction phase, we initiate our processing with two distinct audio sequences: the first contains mixed audio, which encompasses the speech of the target speaker as well as interfering speakers, while the second features reference audio, representing another speech segment from the same target speaker. These two distinct audio sequences yield magnitude spectrograms referred to as $X_{i}$ and $X_{r}$.

Our study adopted the pre-trained VoiceFilter model \cite{wang2018voicefilter}, a specialized speaker extraction network recognized for its capacity to generate speaker-dependent masks. These masks act as effective filters, selectively isolating the target speaker's speech while suppressing interference from non-target speakers. Both $X_{i}$ and $X_{r}$ serve as input to the VoiceFilter to generate a speaker-dependent mask, which is the product of a fusion of information from both the mixed audio and the reference audio. This strategic integration equips the mask to intelligently combat the disruptive influence of interfering speakers, thereby significantly enhancing the model's resilience. Subsequently, this speaker-dependent mask is applied to $X_{i}$, resulting in the creation of a masked spectrogram denoted as $X_{m}$. This mask prediction process can be succinctly represented as follows:

\begin{equation}
    X_{m} = X_{i} \otimes \text{VoiceFilter}(X_{i}, X_{r})
    \label{eq:voicefilter}
\end{equation}

Finally, we concatenate the masked magnitude spectrogram with the phase spectrogram. These features later serve as the input for the DoA estimation phase.

\subsection{DoA Estimation Phase}
In the pursuit of estimating the DoA for the target speaker using the masked spectrogram, we propose a LSTM architecture. This architecture begins with two fully connected layers employing Rectified Linear Unit (ReLU) activation functions \cite{nair2010rectified} and batch normalization \cite{ioffe2015batch} to extract higher-level features from the input. This augmented feature map then feeds into a Bidirectional Gated Recurrent Unit (Bidirectional-GRU) layer, which possesses the capability to capture temporal dependencies and context from both past and future data points \cite{adavanne2018direction}. The output is subsequently directed to two more fully connected layers with a Sigmoid activation function to yield the target DoA estimate.

We adopt the Mean Square Error (MSE) loss for the posterior probability-based coding in Eq.\ref{eq:posterior} to optimize the network,

\begin{equation}
\mathcal{L} = \sum_{t=1}^{T}\sum_{\theta=1}^{360}\left \| p_{t}(\theta) - \Bar{p}_{t}(\theta) \right \|_{2}^{2}
\label{eq:MSE}
\end{equation}

\section{Experiments}
\subsection{Dataset}
We partitioned the VoxCeleb1 \cite{nagrani2017voxceleb} dataset into two non-overlapping subsets: 90\% for training and 10\% for testing. To enhance realism, we incorporated room impulse responses (RIRs) \cite{diaz2021gpurir} to convolve audio segments and create 2-channel mixtures. The specifications of our dataset can be viewed in Tab.1.

\begin{table}[!tb]
\caption{Specifications of our dataset}
\begin{center}
\begin{tabular}{ccc}
\toprule
    \toprule
    \textbf{Voxceleb1} & Training & Testing \\
    \midrule
    \# of audio clips & 4,386 & 488 \\
    \# of distinct speakers & 40 & 40\\
    average duration & 8.27s & 8.29s\\
    \midrule
    \midrule
    \textbf{RIR Files} & Training & Testing\\
    \midrule
    \# of simulated files & 18,020 & 2,000\\
    Range of DoA & $[10.18^{\circ},166.96^{\circ}]$ & $[10.18^{\circ},166.96^{\circ}$]\\
    \bottomrule
    \bottomrule
\end{tabular}
\end{center}
\label{tab:dataset}
\end{table}

For each training sample, we randomly selected two distinct speakers—one serving as the target speaker and the other as the interfering speaker. From the target speaker, we chose two audio segments, with one serving as the reference audio. Additionally, we convolved another audio from the target speaker and the audio from the interfering speaker using two corresponding RIR files. By combining the two convolved audio segments and normalizing the outcome, we generated a comprehensive training sample. 

To further assess the noise robustness of our proposed methods, we evaluated SNRs across a range from -10 $dB$ to 10 $dB$. Here, we considered the audio from the interfering speaker as the noise signal 
$d$, while the audio from the target speaker served as the source signal $s$. Given SNR values, we computed the ratio between the noise signal and the source signal as SNR$= 10 \cdot log_{10}(\frac{\sum_{T}^{t=1}s_{t}^{2}}{\sum_{T}^{t=1}n_{t}^{2}})$.

\begin{table*}[!htbp]
\caption{Experimental results of speaker DoA estimation under different SNRs ($\Delta$ denotes our performance improvement over the second-best method).}
\label{tab_fwsc}
\begin{tabular*}{\linewidth}{*{12}{c}}
\toprule
		\toprule
  \multirow{2}*{\textbf{Model}}& \multicolumn{2}{c}{-10 $dB$} & \multicolumn{2}{c}{-5 $dB$} & \multicolumn{2}{c}{0 $dB$} &
  \multicolumn{2}{c}{5 $dB$} & \multicolumn{2}{c}{10 $dB$}\\
  \cmidrule(lr){2-3}\cmidrule(lr){4-5}\cmidrule(lr){6-7}
  \cmidrule(lr){8-9}\cmidrule(lr){10-11}
  & \small{MAE($^{\circ}$)} & \small{ACC(\%)} & \small{MAE($^{\circ}$)} & \small{ACC(\%)}& \small{MAE($^{\circ}$)} & \small{ACC(\%)} & \small{MAE($^{\circ}$)} & \small{ACC(\%)} & \small{MAE($^{\circ}$)} & \small{ACC(\%)} \\
  \midrule
  MLP-GCC\cite{he2018deep}  & 16.71 & 62.44 & 13.12 &  65.37 & 12.87 & 69.65 & 10.47 & 79.97 & 3.83 & 92.72  \\
  STFT-Resnet\cite{he2021neural} & 12.87 & 65.79 & 12.62 & 70.38 & 11.23 & 78.44 & 8.65 & 83.67 & 2.08 & 95.18 \\
  CRNN\cite{cao2019polyphonic} & 13.14 & 70.63 & 11.88 & 75.71 & 9.45 & 82.13 & 4.28 & 87.32 & 1.83 & 96.48\\
  LSTM & 12.70 & 53.26  & 14.20 & 58.79 & 13.56 & 70.07 & 6.98 & 85.47 & 1.51 & 96.92 \\
  LocSelect (ours) & \textbf{3.55} & \textbf{87.40} & \textbf{3.47} & \textbf{89.25} & \textbf{4.34} & \textbf{91.24} & \textbf{1.92} & \textbf{95.46} & \textbf{0.92} & \textbf{97.36} \\
  \hdashline
  $\Delta$ & -9.15 & +16.77 & -8.41 & +13.54 & -5.11 & +9.11 & -2.36 & +8.14 & -0.59 & +0.44\\
  \bottomrule
\bottomrule
\end{tabular*}
\end{table*}

\begin{figure*}[htb] 

\begin{minipage}[b]{0.18\linewidth} 
  \centering
  \centerline{\includegraphics[width=4.5cm]{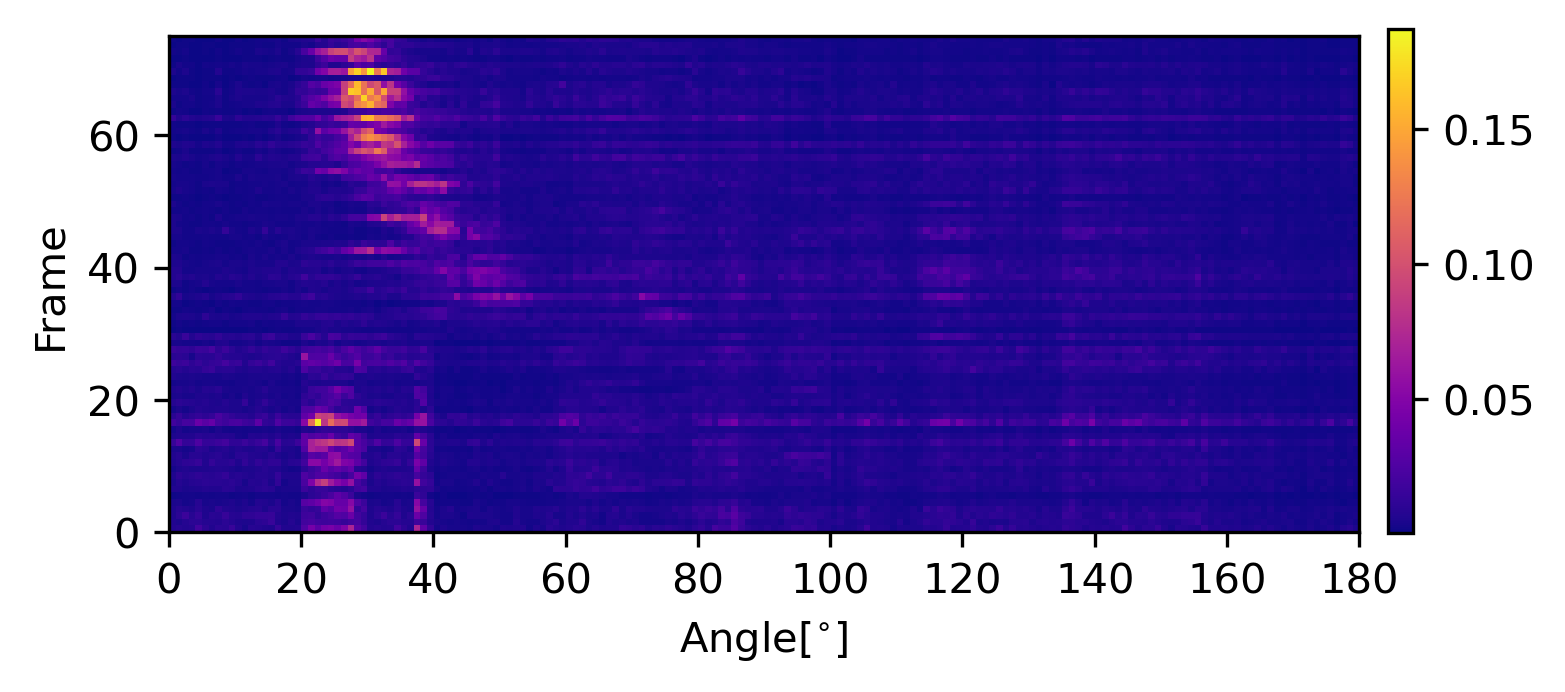}}
  \centerline{\small (a) MLP-GCC}\medskip
\end{minipage}
\begin{minipage}[b]{.18\linewidth}
  \centering
  \centerline{\includegraphics[width=4.5cm]{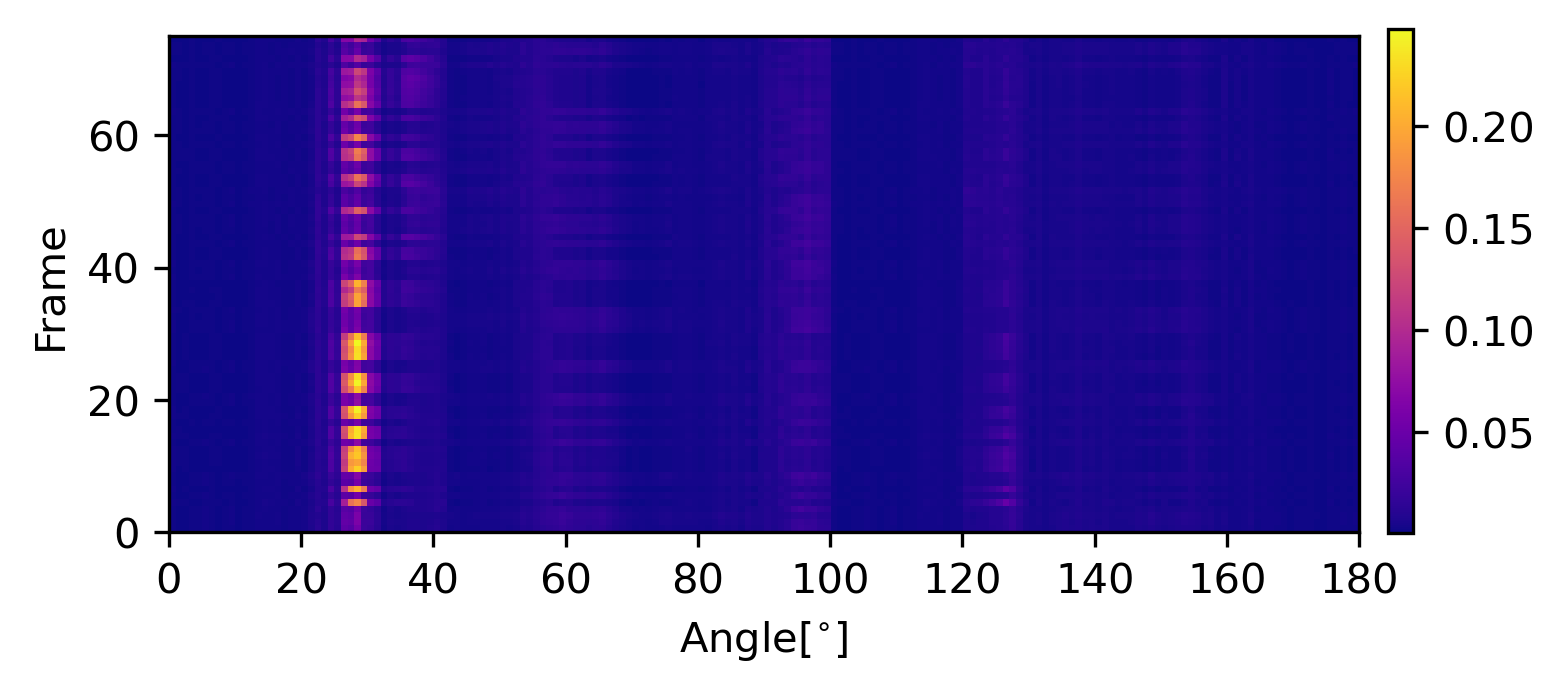}}
  \centerline{\small (b) STFT-Resnet}\medskip
\end{minipage}
\hfill              
\begin{minipage}[b]{0.18\linewidth}
  \centering
  \centerline{\includegraphics[width=4.5cm]{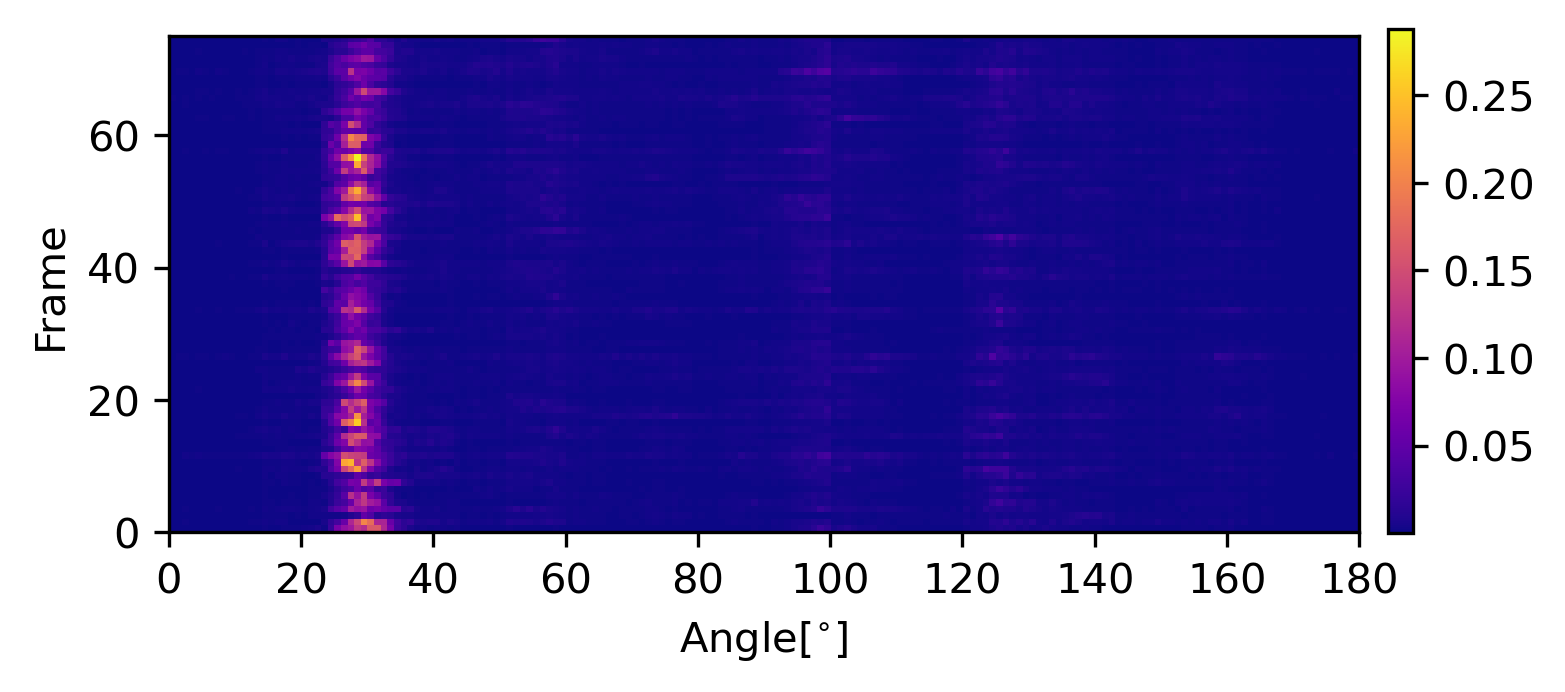}}
  \centerline{\small(c) CRNN}\medskip
\end{minipage}
\hfill
\begin{minipage}[b]{0.18\linewidth}
  \centering
  \centerline{\includegraphics[width=4.5cm]{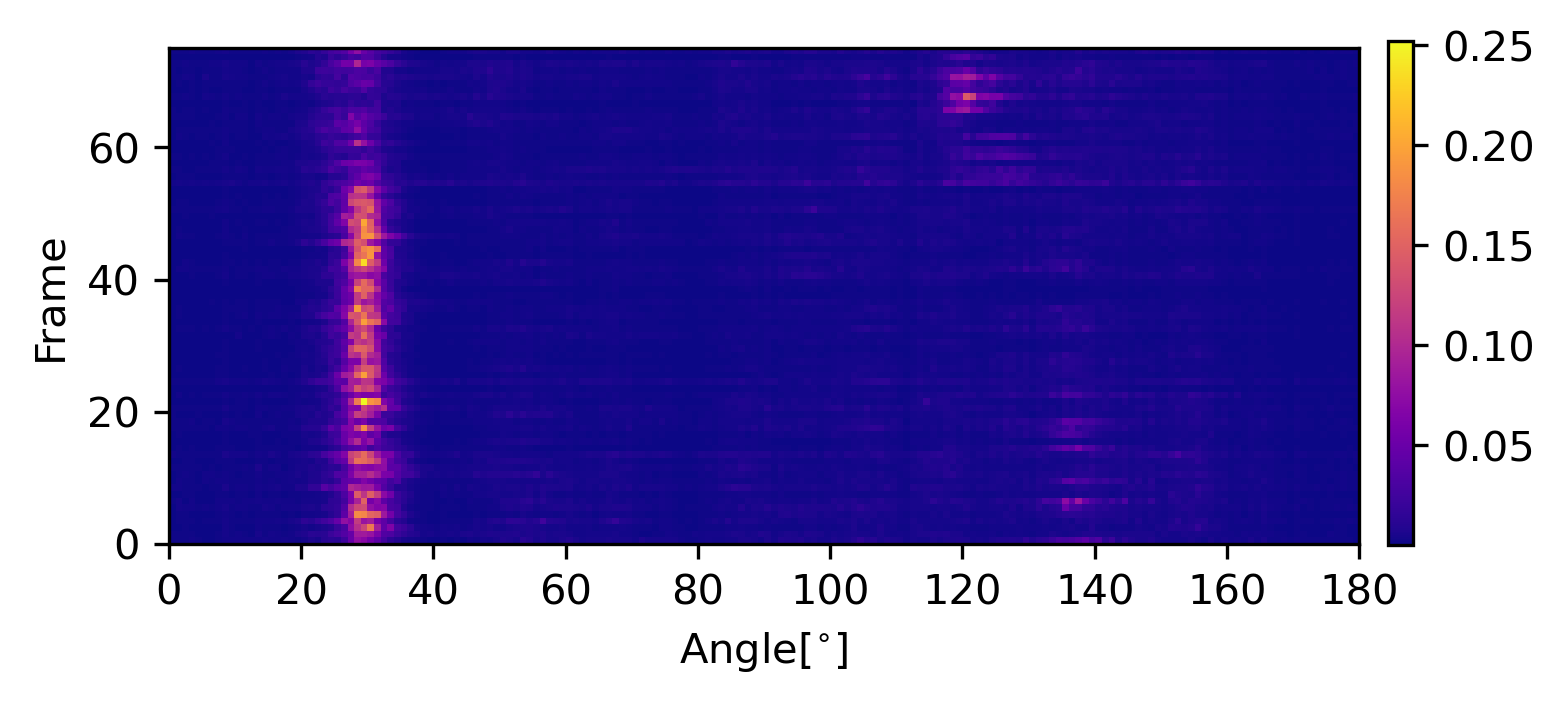}}
  \centerline{\small(d) LSTM}\medskip
\end{minipage}
\hfill
\begin{minipage}[b]{0.18\linewidth}
  \centering
  \centerline{\includegraphics[width=4.5cm]{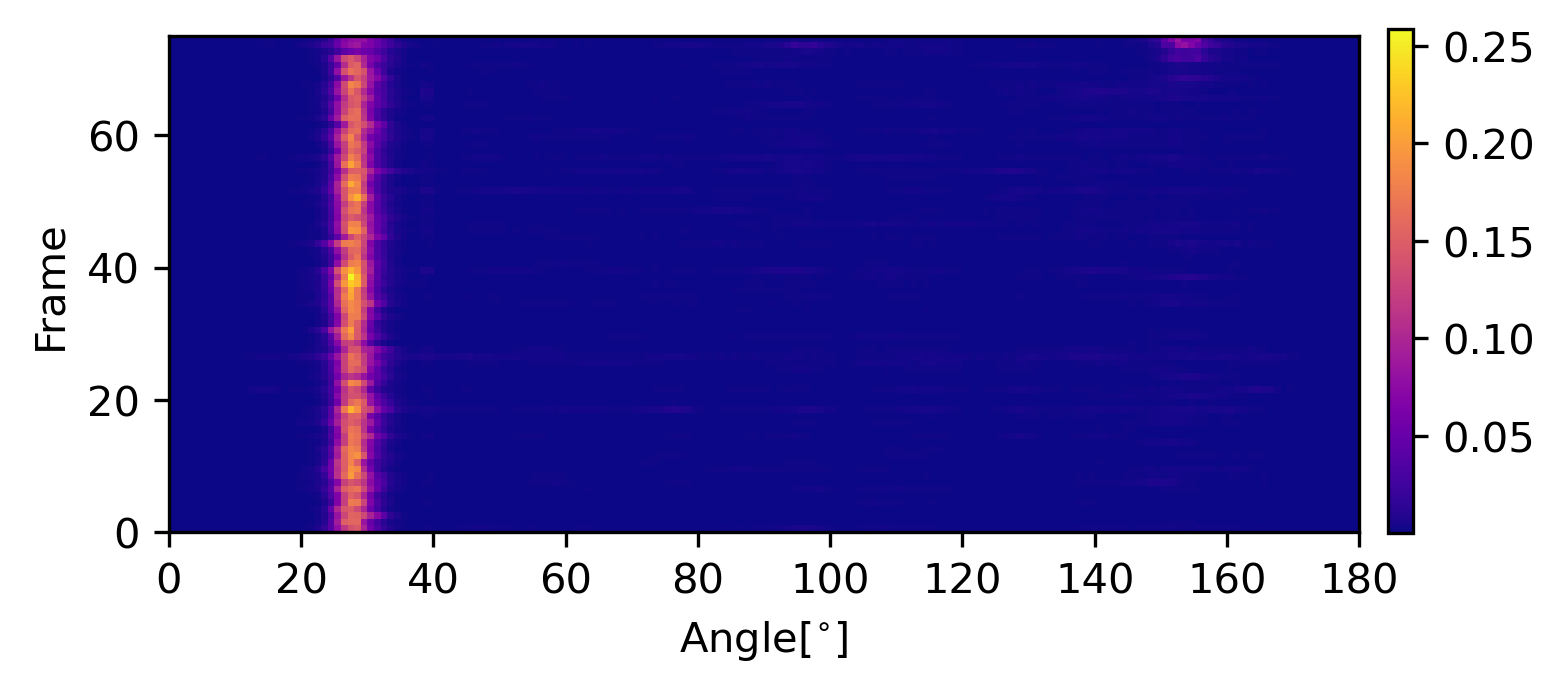}}
  \centerline{\small (e) LocSelect (ours)}\medskip
\end{minipage}
\vspace{-2mm}
\caption{The network outputs (pre-Sigmoid) of a from different models. A single sample is randomly chosen at SNR = 0 $dB$, and each of the models uses this sample to estimate the DoA for the target speaker. The vertical and horizontal axes represent time frames and DoA, respectively. In this specific example, the target speaker is located at $28^{\circ}$, whereas the interfering speaker is  at $122^{\circ}$.}
\label{fig:res}
\end{figure*}
\subsection{Performance Metrics}
We evaluate the methods using MAE and ACC (`$\downarrow$' and `$\uparrow$' indicate the higher and the lower value corresponds to the better results). Specifically, MAE is measured as the average distance between the ground truth DoA $\theta$ and the estimated DoA $\Bar{\theta}$ of the target speaker,
\begin{equation}
    \text{MAE}(\downarrow) = \frac{1}{T}\sum_{t=1}^{T}\|\theta_{t} - \Bar{\theta_{t}}\|
    \label{eq:MAE}
\end{equation}
ACC denotes the ratio of correct estimates at all frames,
\begin{equation}
    \text{ACC}(\uparrow) = \frac{1}{T}\sum_{t=1}^{T} \sigma_{t}, \sigma_{t} =
\left\{  
     \begin{array}{lr}  
     1, \ \|\theta_{t} - \Bar{\theta_{t}}\| \leq \rho&  \\  
     0 &
     \end{array}  
\right.  
\label{eq:ACC}
\end{equation}
where $\rho$ is a tolerance which is set to $5^{\circ}$ in this work.
\subsection{Baselines}
We compared our proposed methods with competitive SOTA methods. Three of them (MLP-GCC, STFT-Resnet, CRNN) \cite{he2018deep,he2021neural} estimate the target speaker's DoA without using reference audio. Moreover, we regarded the LSTM module without using mask estimation as one another baseline, which can show the effectiveness of using reference audio. In all cases, the network's output is considered as the estimated DoA of the target speaker. We briefly summarize the methodology of the baseline methods as follows:
\begin{enumerate}[itemsep=2pt]
    \item   GCC-MLP \cite{he2018deep}: It is a well-known multi-speaker localization method that serves as the building block of different methods~\cite{wang2021gcc}. Specifically, it integrates GCC-PHAT as the input acoustic feature and employs an MLP network with three fully-connected hidden layers for DoA classification.
    \item STFT-ResNet \cite{he2021neural}: It uses STFT as input where the involved residual connections \cite{he2016deep} enhance high-level feature extraction, and the two-stage training strategy prevents the network from getting stuck in local optima.
    \item CRNN \cite{cao2019polyphonic}: It stacks GCC-PHAT and log-mel spectrogram as the inputs to a CRNN network. Although framework is originally designed as a two-stage method for sound event detection and localization, we only used its DoA estimation branch to determine the DoA for the target speaker.
    \item LSTM: Our proposal without VoiceFilter. It uses STFT as input, employing bidirectional GRU to better understand contextual information within audio sequences to learn the DoA feature representation.
\end{enumerate}
\subsection{Parameter settings}
All the methods were tested with the same parameter settings for fair comparisons. Specifically, audio signals were sampled at 16 kHz which were then segmented to clips of 5 seconds.
STFT was computed with a 400-point Hanning window at a hop size of 160 points.We trained all the models for 30 epochs with a batch size of 8 and a learning rate of 0.0001. We use Adam as the network Optimizer \cite{kingma2014adam}. 
\subsection{Results}
Tab.2 presents the results of our experiments conducted on the Voxceleb1 test set. The outcomes are categorized and reported separately for various subsets, each with a known number of speakers. Remarkably, the top-performing results in each column are highlighted in bold. The findings in Tab.2 underscore the efficacy of the speaker-dependent mask. In particular, our proposed LocSelect significantly reduces the MAE from 13.56$^{\circ}$ (LSTM) to 4.34$^{\circ}$, alongside an increase in accuracy from 70.07\% to 91.24\%. These improvements firmly underline the advantages of the speaker-dependent mask. Compared to other baseline methods for DoA estimation, such as GCC-MLP, STFT-Resnet and CRNN, our proposed methods consistently exhibit superior results, featuring lower MAE values and higher accuracy rates.

To offer a visual representation of the localization performance among different methods, we randomly select a sample and plot the network outputs (pre-Sigmoid) across time frames in Fig.3. Specifically, in Fig.3 (a) and Fig.3 (d), it's evident that the absence of a speaker-dependent mask adversely affects both MLP-GCC and LSTM, leading to distortion in the direction of the target speaker due to interference. Although STFT-Resnet (Fig.3 (b)) and CRNN (Fif.3(c)) perform better than their predecessors, they still struggle with maintaining output continuity. In contrast, as demonstrated in Fig.3 (e), the inclusion of the speaker-dependent mask significantly mitigates the influence of interfering speakers. This allows the models to concentrate on precise target speaker localization, further highlighting the effectiveness of our approach.

Subsequently, we conducted a rigorous assessment of our proposed networks' robustness. We introduced variations in the ratio between target speaker audio and interfering speaker audio in mixed audios, aligning with SNR values ranging from -10 $dB$ to 10 $dB$. As the SNR drops below 0, our proposed LocSelect maintained exceptional performance, characterized by an MAE less than $5^{\circ}$ and an ACC exceeding 80\%. In contrast, other baseline methods exhibited declining accuracy below 80\% as the SNR degraded. Notably, at an SNR of -10 $dB$, our proposed LocSelect achieved an accuracy of 87.40\%, surpassing the LSTM approach's performance, which recorded an accuracy of 53.26\%. This improvement can be attributed to the speaker-dependent mask generated in LocSelect, which incorporates information from both reference audio and mixed audio. This mask captures a comprehensive representation of the target speaker, thereby bolstering the system's resilience in challenging conditions.
\section{Conclusion}
This paper presents a novel target speaker localization algorithm utilizing a selective hearing mechanism, where a reference speech from the target speaker generates a speaker-dependent mask for exclusive target speaker localization among multiple speakers. Evaluation results demonstrate the superiority of our approach, achieving a notable 16.77\% increase in ACC and a substantial 9.15$^{\circ}$ reduction in MAE compared to SOTA methods when SNR = -10 $dB$.

Our future research will tackle
more complex scenarios, including ambient noise, unknown and varying number of interfering speakers. Another intriguing avenue of investigation is training our system to extract the DoA for multiple target speakers simultaneously. Rather than focusing on a single target speaker, we intend to leverage reference audios from several target speakers to enable the concurrent extraction of DoA for multiple sources.

\newpage
\ninept
\bibliographystyle{IEEEbib}
\bibliography{strings,refs}

\begin{thebibliography}{10}

\bibitem{1162830}
C.~Knapp and G.~Carter,
\newblock ``The generalized correlation method for estimation of time delay,''
\newblock {\em IEEE/ACM Trans. Audio, Speech, Lang. Process.}, vol. 24, no. 4, pp. 320--327, Aug 1976.

\bibitem{1143830}
R.~Schmidt,
\newblock ``Multiple emitter location and signal parameter estimation,''
\newblock {\em IEEE Trans. on Antennas and Propagation}, vol. 34, no. 3, pp. 276--280, 1986.

\bibitem{takeda2018unsupervised}
R.~Takeda, Y.~Kudo, K.~Takashima, Y.~Kitamura, and K.~Komatani,
\newblock ``Unsupervised adaptation of neural networks for discriminative sound source localization with eliminative constraint,''
\newblock in {\em Proc. ICASSP}, 2018.

\bibitem{7953333}
P.~Pertilä and E.~Cakir,
\newblock ``Robust direction estimation with convolutional neural networks based steered response power,''
\newblock in {\em Proc. ICASSP}, 2017.

\bibitem{xue2019direct}
W.~Xue, Y.~Tong, G.~Ding, C.~Zhang, T.~Ma, X.~He, and B.~Zhou,
\newblock ``Direct-path signal cross-correlation estimation for sound source localization in reverberation.,''
\newblock in {\em Proc. Interspeech}, 2019.

\bibitem{wang2018robust}
Z.-Q. Wang, X.~Zhang, and D.~Wang,
\newblock ``Robust {TDOA} estimation based on time-frequency masking and deep neural networks.,''
\newblock in {\em Proc. Interspeech}, 2018.

\bibitem{shmuel2023deep}
D.~H. Shmuel, J.~P. Merkofer, G.~Revach, R.~J. van Sloun, and N.~Shlezinger,
\newblock ``Deep root music algorithm for data-driven doa estimation,''
\newblock in {\em Proc. ICASSP}, 2023.

\bibitem{9564893}
C.~Liu, W.~Feng, H.~Li, and H.~Zhu,
\newblock ``Single snapshot doa estimation based on spatial smoothing music and cnn,''
\newblock in {\em Proc. ICSPCC}, 2021.

\bibitem{chakrabarty2017broadband}
S.~Chakrabarty and E.~A. Habets,
\newblock ``Broadband {DOA} estimation using convolutional neural networks trained with noise signals,''
\newblock in {\em Proc. WASPAA}. IEEE, 2017.

\bibitem{perotin2018crnn}
L.~Perotin, R.~Serizel, E.~Vincent, and A.~Gu{\'e}rin,
\newblock ``{CRNN}-based joint azimuth and elevation localization with the ambisonics intensity vector,''
\newblock in {\em Proc. IWAENC}. IEEE, 2018.

\bibitem{cao2019polyphonic}
Y.~Cao, Q.~Kong, T.~Iqbal, F.~An, W.~Wang, and M.~D. Plumbley,
\newblock ``Polyphonic sound event detection and localization using a two-stage strategy,''
\newblock {\em arXiv preprint arXiv:1905.00268}, 2019.

\bibitem{nguyen2021general}
T.~N.~T. Nguyen, N.~K. Nguyen, H.~Phan, L.~Pham, K.~Ooi, D.~L. Jones, and W.-S. Gan,
\newblock ``A general network architecture for sound event localization and detection using transfer learning and recurrent neural network,''
\newblock in {\em Proc. ICASSP}, 2021.

\bibitem{sivasankaran2018keyword}
S.~Sivasankaran, E.~Vincent, and D.~Fohr,
\newblock ``Keyword-based speaker localization: Localizing a target speaker in a multi-speaker environment,''
\newblock in {\em Proc. Interspeech}, 2018.

\bibitem{wang2018target}
Z.~Wang, J.~Li, and Y.~Yan,
\newblock ``Target speaker localization based on the complex watson mixture model and time-frequency selection neural network,''
\newblock {\em Applied Sciences}, vol. 8, no. 11, pp. 2326, 2018.

\bibitem{4490206}
R.~Munkong and B.-H. Juang,
\newblock ``Auditory perception and cognition,''
\newblock {\em IEEE Signal Processing Magazine}, vol. 25, no. 3, pp. 98--117, 2008.

\bibitem{wang2018voicefilter}
Q.~Wang, H.~Muckenhirn, K.~Wilson, P.~Sridhar, Z.~Wu, J.~Hershey, R.~A. Saurous, R.~J. Weiss, Y.~Jia, and I.~L. Moreno,
\newblock ``Voicefilter: Targeted voice separation by speaker-conditioned spectrogram masking,''
\newblock {\em arXiv preprint arXiv:1810.04826}, 2018.

\bibitem{he2018deep}
W.~He, P.~Motlicek, and J.-M. Odobez,
\newblock ``Deep neural networks for multiple speaker detection and localization,''
\newblock in {\em Proc. ICRA}, 2018.

\bibitem{he2021neural}
W.~He, P.~Motlicek, and J.-M. Odobez,
\newblock ``Neural network adaptation and data augmentation for multi-speaker direction-of-arrival estimation,''
\newblock {\em IEEE/ACM Trans. on Audio, Speech, and Language Processing}, vol. 29, pp. 1303--1317, 2021.

\bibitem{wang2020ustc}
Q.~Wang, H.~Wu, Z.~Jing, F.~Ma, Y.~Fang, Y.~Wang, T.~Chen, J.~Pan, J.~Du, and C.-H. Lee,
\newblock ``The {USTC-iFlytek} system for sound event localization and detection of {DCASE2020} challenge,''
\newblock {\em IEEE AASP Chall. Detect. Classif. Acoust. Scenes Events}, 2020.

\bibitem{nguyen2020ensemble}
T.~N.~T. Nguyen, D.~L. Jones, and W.-S. Gan,
\newblock ``Ensemble of sequence matching networks for dynamic sound event localization, detection, and tracking,''
\newblock {\em IEEE AASP Chall. Detect. Classif. Acoust. Scenes Events}, pp. 120--124, 2020.

\bibitem{nair2010rectified}
V.~Nair and G.~E. Hinton,
\newblock ``Rectified linear units improve restricted boltzmann machines,''
\newblock in {\em Proc. ICML}, 2010.

\bibitem{ioffe2015batch}
S.~Ioffe and C.~Szegedy,
\newblock ``Batch normalization: Accelerating deep network training by reducing internal covariate shift,''
\newblock in {\em Proc. ICML}, 2015.

\bibitem{adavanne2018direction}
S.~Adavanne, A.~Politis, and T.~Virtanen,
\newblock ``Direction of arrival estimation for multiple sound sources using convolutional recurrent neural network,''
\newblock in {\em Proc. EUSIPCO}, 2018.

\bibitem{nagrani2017voxceleb}
A.~Nagrani, J.~S. Chung, and A.~Zisserman,
\newblock ``{VoxCeleb}: a large-scale speaker identification dataset,''
\newblock {\em arXiv preprint arXiv:1706.08612}, 2017.

\bibitem{diaz2021gpurir}
D.~Diaz-Guerra, A.~Miguel, and J.~R. Beltran,
\newblock ``{gpuRIR}: A python library for room impulse response simulation with gpu acceleration,''
\newblock {\em Multimedia Tools and Applications}, vol. 80, pp. 5653--5671, 2021.

\bibitem{wang2021gcc}
J.~Wang, X.~Qian, Z.~Pan, M.~Zhang, and H.~Li,
\newblock ``{GCC-PHAT} with speech-oriented attention for robotic sound source localization,''
\newblock in {\em Proc. ICRA}, 2021.

\bibitem{he2016deep}
K.~He, X.~Zhang, S.~Ren, and J.~Sun,
\newblock ``Deep residual learning for image recognition,''
\newblock in {\em Proc. CVPR}, 2016.

\bibitem{kingma2014adam}
D.~P. Kingma and J.~Ba,
\newblock ``Adam: A method for stochastic optimization,''
\newblock {\em arXiv preprint arXiv:1412.6980}, 2014.

\end{thebibliography}

\end{document}